# Title: Defect Level Switching for Highly-Nonlinear and Hysteretic Electronic Devices

**Authors:** Han Yin[1], Abinash Kumar[2], James M. LeBeau[2], R. Jaramillo[2*]

**Affiliations:**

[1]Department of Mechanical Engineering, Massachusetts Institute of Technology, Cambrige, MA 02139, USA

[2]Department of Materials Science & Engineering, Massachusetts Institute of Technology, Cambridge, MA 02139, USA

[*]Corresponding author. Email: rjaramil@mit.edu

**Abstract:**

Nonlinear and hysteretic electrical devices are needed for applications from circuit protection to next-generation computing. Widely-studied devices for resistive switching are based on mass transport, such as the drift of ions in an electric field, and on collective phenomena, such as insulator-metal transitions. We ask whether the large photoconductive response known in many semiconductors can be stimulated in the dark and harnessed to design electrical devices. We design and test devices based on photoconductive CdS, and our results are consistent with the hypothesis that resistive switching arises from point defects that switch between deep- and shallow-donor configurations: defect level switching (DLS). This new electronic device design principle - *photoconductivity without photons* – leverages decades of research on photoconductivity and defect spectroscopy. It is easily generalized and will enable the rational design of new nonlinear, hysteretic devices for future electronics.

**Main Text:**

Two terminal resistive devices are basic building blocks for electronics. Simple resistors obey Ohm's law and exhibit a linear relationship between current (*I*) and voltage (*V*). Semiconductor devices often exhibit nonlinear *I-V* characteristics, such as diode behavior. More complicated devices exhibit *I-V* behavior that is both nonlinear and hysteretic. Nonlinear, hysteretic behavior is needed for many applications including circuit protection (*e.g.* fuses), oscillators (*e.g.* Gunn diode), memory selectors (*e.g.* ovonic switches), and emerging concepts for computing (*e.g.* resistive switches) [1–4]. Research on resistive switching has expanded rapidly as future computing applications come into view including storage-class memory, analog and resistive computing, compute-in-memory, and neuromorphic computing [5–10]. Charge-based resistive switching has been observed in many materials, and a taxonomy has emerged to describe the different underlying mechanisms [8,10]. Most resistive switches are based on the mass transport of atoms and ions, which move in response to electrochemical forces. Other categories of charge-based resistive switching are based on collective phenomena, such as amorphous-crystalline transformations (*e.g.* phase-change materials) and insulator-metal transitions driven by electron-electron correlation.

Here we introduce a mechanism of charge-based resistive switching that does not require mass transport or collective phenomena. We call it defect level switching (DLS) because it is based on atomic structure changes around semiconductor point defects; these changes affect the transition energy levels, and thereby the electrical conductivity. The underlying mechanism is grounded in well-established semiconductor physics, and is already known in a number of different material systems. DLS therefore presents a framework for rational design of nonlinear, hysteretic electronic devices with properties (*e.g.* switching speed and energy, retention time, polarity) that can be accurately predicted by theory and simulation. We illustrate this predictive design framework by describing and then experimentally demonstrating two, quite different resistive switches – a DLS device that is in a high-conductivity ("ON") state as-fabricated and at equilibrium (like a Gunn diode), and a more traditional interface switching device that is in a low-conductivity ("OFF") state as-fabricated and at equilibrium – that both rely on the same active material.

Point defects in semiconductors are characterized by their transition energy levels, the thermodynamic potentials at which their equilibrium ionization state changes. In some cases, point defects can feature multiple transition levels, widely separated in energy, due to the combined effects of charge-lattice coupling and electron-electron correlation. Such systems are often bistable, with two distinct atomic arrangements corresponding to deep- and shallow-donor states (for *n*-type; acceptor for *p*-type), where states' energy depend on ionization; see **Fig. 1**. These cases of strong correlation violate the independent quasiparticle assumption that underlies semiconductor electronic structure models; this is unsurprising for localized defects, whose behavior often more closely resembles molecules than extended solids. Bistable point defects are responsible for large and persistent photoconductivity in well-studied semiconductor materials including III-Vs (*e.g.* AlGaAs, GaInNAs), II-VIs (*e.g.* ZnO, CdS), and chalcopyrites (*e.g.* $CuInS_2$, $CuGaSe_2$) [11–14].

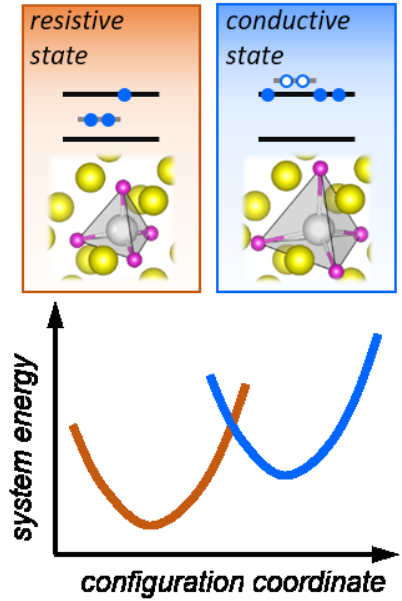

**Figure 1.** Defect level switching (DLS). Configuration coordinate diagram illustrated here for CdS, for which the double-ionization of neutral sulfur vacancies induces a lattice relaxation into a metastable and highly-conductive state, resulting in a billion-fold increase in conductivity upon white light exposure [14].

We invite the question of whether this type and magnitude of response could be controlled in an electrical device in the dark: akin to photoconductivity, but without photons. To achieve this functionality, we need to re-create the minority carrier injection and defect ionization as occurs in photoconductivity. We use photoconductive CdS thin films as our active material, and semiconductor device simulations (*i.e.* TCAD software) as our design tool. For a device to be in the ON state as-fabricated and at equilibrium, we require that at zero bias the Fermi energy ($E_F$) crosses the transition level of a DLS-active deep donor, resulting in level switching into a shallow donor state and the creation of a high-conductivity sheath. The device will show resistive switching into an OFF state under forward bias if the bands bend so as to reduce the width of this high-conductivity sheath.

In **Fig. 2a** we show the design for the equilibrium-ON device. We use $MoO_3$ as a hole-injection layer [15,16]. Simulations predict that this $MoO_3$/CdS device will switch into an OFF state under forward bias, as the high-conductivity sheath of ionized sulfur vacancies narrows. This is remarkable because we predict that *n*-type CdS will become *less* conductive upon electron injection. In **Fig. 2b** we show the results of our experimental device tests. The behavior is as-predicted for the DLS hypothesis and device design. The device as-fabricated at equilibrium is in an ON state, and the *I-V* curve resembles a simple resistor. Under forward bias the device abruptly switches into a metastable OFF state, with a rectifying and diode-like *I-V* curve. The original ON state can be recovered by sweeping into reverse bias, like a bipolar resistive switch.

The equilibrium-ON device demonstrates bipolar switching, but the switching polarity and further hypothesis tests (below) distinguish it from bipolar resistive switches based on defect drift motion (*i.e.* interface switching) [10]. The peculiar behavior of our equilibrium-ON device is more similar to a Gunn diode. Both our devices and Gunn diodes are distinguished by a conductivity drop upon majority carrier injection, although the underlying mechanisms are different: carrier re-capture at point defects in the DLS device *vs.* injection into lower mobility conduction band minima in a Gunn diode. Equilibrium-ON DLS devices may be useful for circuit protection (*i.e.* like a breaker), or as oscillators.

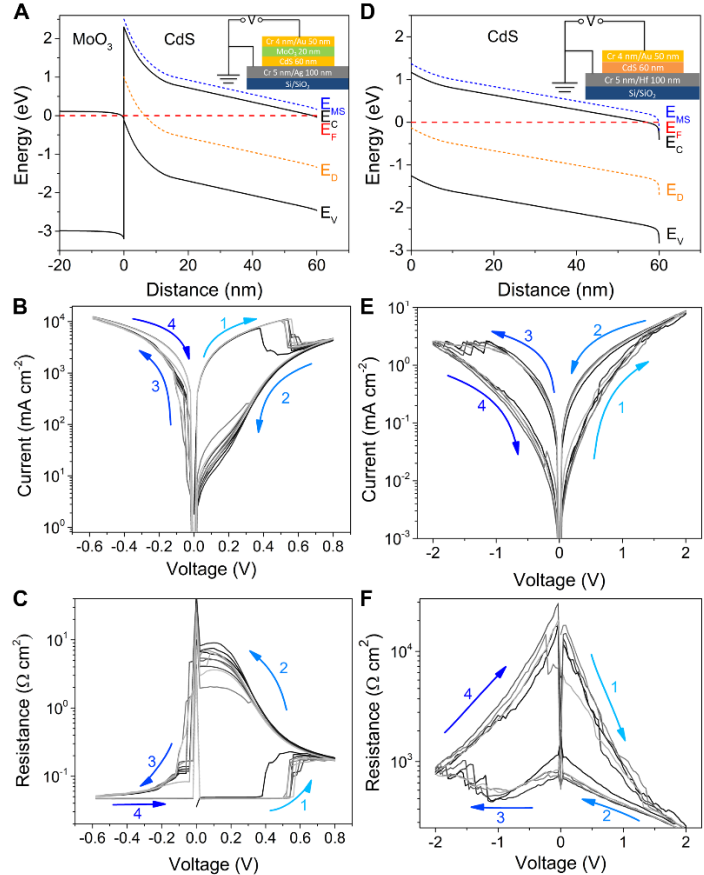

**Fig. 2.** Predictive design and experimental tests of DLS devices. Simulations show the device band diagrams at zero bias; experimental tests are on 1 mm² devices. $E_C$ = conduction band edge, $E_V$ = valence band edge, $E_F$ = Fermi energy, $E_D$ = transition energy of neutral DLS-active defect, $E_{MS}$ = transition energy of DLS-active defect in the ionized and metastable state. (**A**) Design of equilibrium-ON device with a hole-injecting contact. (**B-C**) Experimental tests of equilibrium-ON device. (**D**) Design of equilibrium-OFF device with inert and non-hole injecting contacts. (**E-F**) Experimental tests of equilibrium-OFF device.

By selecting non-hole injecting contacts we can also design an equilibrium-OFF device. In **Fig. 2d** we show the design for an equilibrium-OFF device with electron-injecting Au and Hf contacts. The Au contact has a thin Cr layer that improves adhesion and suppresses possible Au migration (*c.f.* **Fig. S3**). Our design simulations predict that the DLS-active deep levels remain neutral, at equilibrium and for all reasonable applied voltages; therefore the device should be in a low-conductivity state as-fabricated, and we expect resistance switching to be controlled by defect motion at the rectifying Au/CdS junction. In **Fig. 4d** we show the results of our experimental device tests. The device is in a low-conductivity state at equilibrium, as predicted, and it exhibits bipolar switching between OFF and ON states. The voltage required for switching is nearly five times larger than for the equilibrium-ON DLS device, and the switching polarity is consistent with mobile ionized donors that modulate the Au/CdS Schottky junction, as is widely-observed in oxides [10].

The results in **Fig. 2** illustrate the predictive design made possible by the DLS mechanism. We also report a number of complementary experiments with equilibrium-ON devices that further test the DLS hypothesis and that fall into two categories: materials processing and substitution, and advanced characterization. The most straightforward tests are those involving material omission and substitution. In **Fig. 3a** we show *I-V* data for devices with the $MoO_3$ or the CdS layer omitted. The resulting Ag/CdS/Au and Ag/$MoO_3$/Au devices are simple resistors and show no sign of resistive switching, thereby demonstrating that the $MoO_3$/CdS heterojunction is paramount. In **Fig. 3b** we demonstrate the joint importance of CdS photoconductivity and the $MoO_3$/CdS heterojunction. A device made using non-photoconductive (NPC) CdS is rectifying and shows no switching, even for applied bias beyond ±2 V. This demonstrates the importance of CdS photoconductivity, which derives from sulfur vacancies [14]. Meanwhile, a device made from photoconductive CdS but with a typical transparent conducting oxide (ITO) instead of $MoO_3$ is weakly rectifying and show no resistive switching. This demonstrates the importance of the $MoO_3$/CdS heterojunction to imitate photoconductivity by injecting minority carriers. $MoO_3$ and ITO are both significantly more conductive than CdS, and therefore the electric field inside the CdS active layer is similar during *I-V* testing for both devices. The absence of switching for the ITO/CdS device demonstrates that, although sulfur vacancies are necessary for the DLS phenomenon, their drift motion does not seem to be essential for switching. Conventional resistive switching is based on the formation of metal filaments in a dielectric matrix, and Ag is a widely-used filament-former [8,17]. We test for Ag filament formation by replacing Ag by relatively inert and immobile Cr. In **Fig. 3c** we show the performance of the resulting Cr/CdS/$MoO_3$ device that shows the same equilibrium-ON switching behavior as the baseline Ag/CdS/$MoO_3$ device. The higher bias required for switching the Cr-based device, and the lower current and relative switching magnitude (e.g. $I_{ON}/I_{OFF}$), result from the substantial series resistance contributed by the Cr bottom contact, Cr being much less conductive than Ag. By way of contrast, when we replace Ag by Au – which is mobile and has a chemical affinity for sulfur – the resulting device performance is non-repeatable and suggestive of both DLS and filament formation mechanisms playing a role (Supplementary Materials). The results that we present in **Fig. 3a-c** for alternative devices related to the baseline device by material substitution or omission (Ag/CdS/Au, Ag/$MoO_3$/Au, Ag/NPC CdS/$MoO_3$, Ag/CdS/ITO, Cr/CdS/$MoO_3$) are all consistent with the DLS hypothesis.

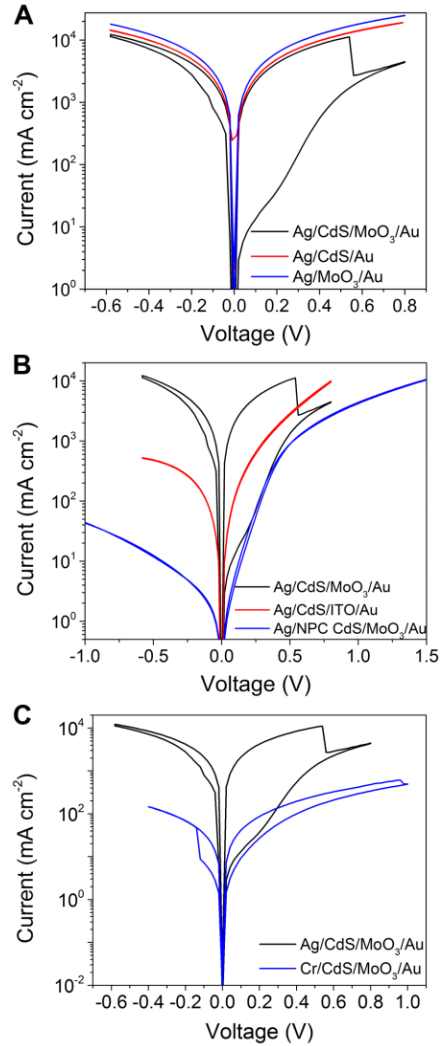

**Figure 3:** Testing the DLS hypothesis by material substitution and omission. (**A**) Ag/CdS/Au and Ag/MoO$_3$/Au devices without the CdS/MoO$_3$ junction are simple resistors and show no switching. (**B**) No switching is observed for devices made using non-photoconductive (NPC) CdS, or substituting ITO in place of MoO$_3$. (**C**) Equilibrium-ON switching is observed for a device with Cr instead of Ag as the back contact; the quantitative differences between the Cr- and Ag-based device *I-V* curves are due to the substantial series resistance contributed by Cr.

We further test the DLS mechanism by advanced characterization of equilibrium-ON devices. In **Fig. 4a** we present a study of the retention kinetics. The DLS mechanism is based on the same atomic mechanism as photoconductivity, and we therefore expect that the kinetics of relaxation to equilibrium would be similar. We find that the time required to return to the equilibrium-ON state after a switch to the OFF state at forward bias varies with temperature as a thermally-activated process, with activation energy $E_{A,DLS} = 0.57 \pm 0.02$ eV. This matches the activation energy for recombination $E_{A,PC} = 0.55 \pm 0.02$ eV that we previously found in studies of CdS photoconductivity [14]. This quantitative match further supports the DLS hypothesis of resistive switching based on the same mechanism as photoconductivity.

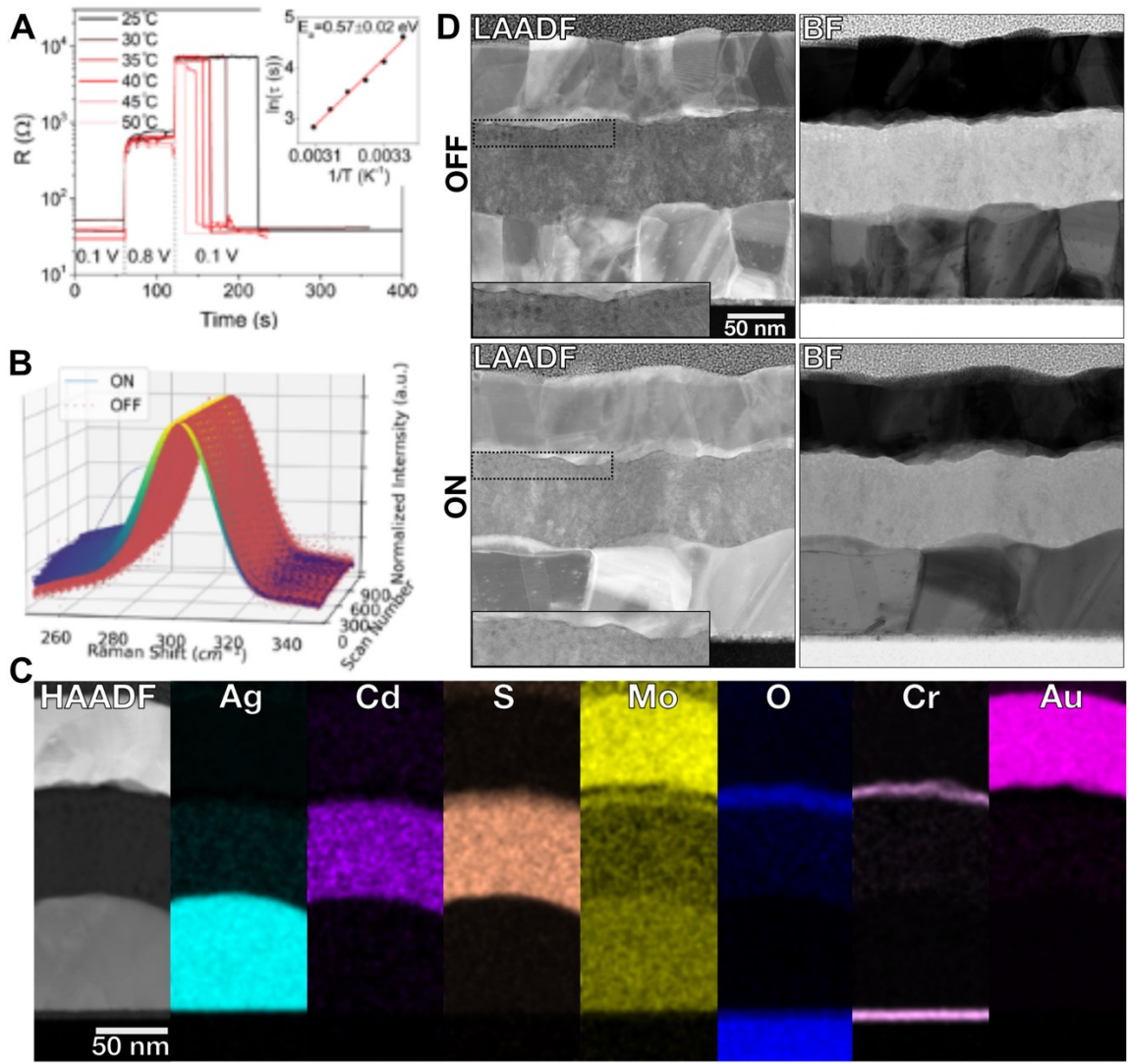

**Figure 4:** Testing the DLS hypothesis by advanced characterization of equilibrium-ON devices. (**A**) Retention kinetics show that switching between states at bistable DLS-active defects is a thermally-activated process, with activation energy consistent with that observed for studies of photoconductivity kinetics. Main panel shows resistance as a function of time while switching between ON and OFF states at different temperatures. Inset shows the temperature-dependence of the characteristic return-to-equilibrium time as an Arrhenius plot, with activation energy $E_a$. (**B**) Raman spectra acquired across a 30×30 μm² area of an equilibrium-ON device with spatial resolution of 1 μm² show a spatially-uniform blue-shift of the 1LO peak upon switching from ON to OFF, consistent with electron injection into sulfur vacancies. We plot the Raman intensity as a function of Raman shift and line scan number to best visualize the blue-shift; the scan numbers correspond to 900 locations across the device area. (C) STEM-EDS elemental maps showing the layer structure of a device set to the OFF state. The thin band of Mo is the $MoO_3$ layer; the thick band is an artifact, due to spectral overlap with Au $M\alpha$ emission.

(D) LAADF and BF images of devices set to the OFF state, and cycled from OFF to ON (insets indicate the low-density regions near the MoO$_3$ layer in off state device).

In **Fig. 4b** we show Raman spectra measured across 30×30 μm$^2$ area of an equilibrium-ON device in both the ON and OFF states, with 1 μm$^2$ spatial resolution set by the excitation spot size. To make possible Raman spectromicroscopy, this device was fabricated with a semi-transparent top contact. The excitation laser wavelength is 532 nm, which is just below the band gap of CdS. This optical excitation may perturb the ON and OFF states that everywhere else in this report are controlled and measured all-electrically and in the dark, but we are still able to draw clear conclusions. We find that the 1LO Raman peak blue-shifts uniformly by 2.29 ± 0.74 cm$^{-1}$ upon switching from ON to OFF, with no evidence of localized features (*e.g.* filaments) at the spatial resolution of the experiment. DLS at sulfur vacancies involves a substantial lattice strain, which may be expected to affect the phonon spectra. The blue-shift in the 1LO peak position that we observe upon switching from ON to OFF – *i.e.* electron injection into CdS -- is consistent with a previous study of the effects of charge transfer on CdS Raman spectra, which study includes results for CdS/MoO$_3$ interfaces [16]. The Raman spectromicroscopy data therefore further support the hypothesis that resistive switching results from ionization state transitions at sulfur vacancies, akin to photoconductivity but engineered in an all-electrical device and without photons.

In **Fig. 4c-d**, we show scanning transmission electron microscopy (STEM) data measured from device cross-sections set to the OFF state and cycled between the OFF and ON states. STEM energy-dispersive spectroscopy (EDS) elemental maps show that all layers are coherent with no visible interdiffusion after the OFF set operation (**Fig. 4c**). The low-angle annular dark field (LAADF) and bright field (BF) images show that CdS layer is dense with no evidence of filament formation (**Fig. 4d**). One peculiar observation is the appearance of low-density nodules within the MoO$_3$ layer (insets of **Fig. 4d**). These nodules are more prominent in the OFF state than in the ON state based on the change in image contrast. We speculate that these low-density regions result from electromigration of O$^{2-}$ ions within the MoO$_3$ layer. It has been shown that MoO$_3$ can exhibit interface switching due to O$^{2-}$ ion electromigration [18]. Although a similar phenomenon may be occurring within our devices, we note that it is distinct from the DLS mechanism that governs resistive switching in the CdS active layer. This is simply apparent from the fact that interface resistive switches based on MoO$_3$ are equilibrium-OFF, unlike the equilibrium-ON DLS device shown here. We also recall that no resistive switching is observed in Ag/MoO3/Au devices (**Fig. 2**).

The DLS mechanism provides a paradigm for rational, predictive design of nonlinear, hysteretic resistive devices. It is based on the long-understood phenomenon of ionization state transitions at bistable point defects in semiconductors and adds an extra dimension to point defect engineering, a mainstay of semiconductor physics and device engineering. Material systems known to feature DLS phenomena include oxides (*e.g.* ZnO), chalcogenides (*e.g.* CdS, CuInS$_2$), and III-Vs (*e.g.* AlGaAs, GaInNAs), and include both *n*-type *p*-type semiconductors; in some cases, DLS may be accompanied by atomic diffusion (*e.g.* protons in SrTiO$_3$) [11–14,19]. Other DLS materials may be discovered by an empirical search for materials with large and persistent photoconductivity, or by a heuristic search using density functional theory (DFT) to identify semiconductor point defects with large lattice configurational changes upon ionization [14]. DLS does not rely on mass transport or collective phenomena, which can provide very useful device functionality, but are resistant to predictive design and often come with inherent challenges of stochasticity and fatigue. The design of DLS devices builds on decades of know-how in semiconductor junction and deep-level spectroscopy, and we suggest that key device properties (*e.g.* switching direction, speed, and energy, retention time, polarity) could be reliably predicted using well-established methods for characterizing semiconductor point defects [20].

*Acknowledgments*

We acknowledge the following persons for informative discussions and technical assistance: Dr. Christopher Thompson (University of Delaware); Prof. Marco Nardone (Bowling Green State University); Profs. Nicholas Fang, Prof. Jeehwan Kim, Dr. Zheng Jie Tan, Dr. Ahmed Tiamiyu, and Thomas Defferriere (MIT). Electron microscopy sample preparation was performed at the Center for Nanoscale Systems (CNS), a member of the National Nanotechnology Coordinated Infrastructure Network (NNCI), which is supported by the National Science Foundation under NSF award no. 1541959. CNS is part of Harvard University. This work was supported by the Office of Naval Research MURI through Grant No. N00014-17-1-2661.

***Supplementary Materials***
Materials and Methods
Supplementary Text
Figs. S1 to S3
Table S1

# Supplementary Materials for "*Defect Level Switching for Highly-Nonlinear and Hysteretic Electronic Devices*"

**Authors:** Han Yin[1], Abinash Kumar[2], James M. LeBeau[2], R. Jaramillo[2*]

**Affiliations:**

[1]Department of Mechanical Engineering, Massachusetts Institute of Technology, Cambrige, MA 02139, USA

[2]Department of Materials Science & Engineering, Massachusetts Institute of Technology, Cambridge, MA 02139, USA

[*]Corresponding author. Email: rjaramil@mit.edu

**This PDF file includes:**

    Materials and Methods
    Supplementary Text
    Figs. S1 to S3
    Table S1

## Materials and Methods

*Device simulation*

We simulate the band diagram of our devices using SCAPS software [21]. The CdS layer contains bistable defects with two possible configurations: a deep double-donor, and a shallow double-donor. The distribution across these two configurations is calculated at an initial working voltage, and this distribution in turn affects the band diagram, in a self-consistent solution. We calculate the distribution for the ON state at a voltage of -0.4 V, and for the OFF state at a voltage of 0.4 V. We show in **Table S1** the parameters used.

**Table S1:** Material parameters used to simulate equilibrium-ON DLS devices.

|  | CdS | $MoO_3$ |
|---|---|---|
| Thickness (nm) | 60 | 20 |
| $E_g$ (eV) | 2.42 | 3.1 |
| $\chi$ (eV) | 4.30 | 6.70 |
| $\epsilon_r$ | 9.0 | 4.60 |
| $N_c$ (cm$^{-3}$) | 2.24×10$^{18}$ | 2.0×10$^{19}$ |
| $N_v$ (cm$^{-3}$) | 1.80×10$^{19}$ | 4.0×10$^{19}$ |
| $v_e$ (cm s$^{-1}$) | 2.6×10$^7$ | 1.0×10$^7$ |
| $v_h$ (cm s$^{-1}$) | 1.3×10$^7$ | 1.0×10$^7$ |
| $\mu_e$ (cm$^2$ V$^{-1}$ s$^{-1}$) | 1.57 | 100 |
| $\mu_h$ (cm$^2$ V$^{-1}$ s$^{-1}$) | 1.0 | 100 |
| $N_D$ (cm$^{-3}$) | 1.0×10$^{10}$ | 2.8×10$^{17}$ |
| Defect levels | | |
| Type | +/0 (Gaussian) | |
| $\Delta E_c$ (eV) | 1.4 | |
| Characteristic energy (eV) | 0.2 | |
| $N_t$ (cm$^{-3}$) | 1×10$^{19}$ | |
| Metastable transition levels | | |
| $N_t$ (cm$^{-3}$) | 2×10$^{18}$ | |
| $E_C - E_{TR}$ (eV) | 0.9 | |
| Defect 1 type | +/0 | |
| $\Delta E_{c1}$ (eV) | 1.1 | |
| Defect 2 type | 2+/+ | |
| $\Delta E_{c1}$ (eV) | -0.2 | |

*Device fabrication*

We fabricate devices on silicon wafers with approximately 100 nm-thick oxide layer. For equilibrium-ON devices we first deposit 4 nm of Cr followed by 100 nm of Ag, using thermal evaporation. We then deposit the CdS layer by chemical bath deposition. In a typical procedure, we first dissolve 306 mg Cd(NO$_3$)$_2$ and 76 mg thiourea in 10 mL DI water, respectively. Next, we heat 76 mL DI water to 70 °C and immerse the substrate into the bath vertically. We then add chemicals to the bath in the following order: $Cd(NO_3)_2$ solution, 14 mL ammonium hydroxide, thiourea solution. The chemical bath is continuously stirred by a magnetic stir bar at 300 rpm, and the deposition lasts for 8 min. We rinse the sample with DI water and sonicate it in isopropanol for 5 min. 20 nm $MoO_3$ is then deposited onto the sample using thermal evaporation, followed by a top contact layer of 4 nm Cr/50 nm Au.

We pattern devices using photolithography. We first spin-coat SPR 700 photoresist, and then expose the pattern for 15 s using a photomask. We then develop the pattern in MF CD-26 developer for 1 min. We remove the Au top contact by ion milling, and we remove the $MoO_3$ and CdS by etching in dilute $(NH_4)OH$ and dilute HCl, respectively. We sonicate the sample in acetone to remove the photoresist. Ion milling was found to be necessary to remove the Au top contact because reactive etches (wet and dry) produced unwanted and messy reactions with the underlying layers.

*Device testing*

We carry out electrical measurements using a custom-made probe station. We measure *I-V* data using a Keithley 2400 source meter in a two-point configuration, using BeCu probes to contact the devices. The substrate and top contact are connected to the LO and HIGH terminals, respectively. We use a voltage sweep rate of 0.2 V/s. The sample temperature is controlled using a thermal chuck and is set to 25 ˚C for all measurements, unless stated otherwise. Time series data are collected using the same Keithley 2400 source meter, with an acquisition rate of 1 Hz.

*Electron Microscopy*

Cross-sectional electron microscopy samples of devices in the OFF & cycled states are prepared using a Ga focused ion beam (Thermo Fisher Scientific Helios 660). STEM characterization of devices is performed with a probe corrected Thermo Fisher Scientific Titan G3 60-300 kV operated at 200 kV with a probe semi-convergence angle of 18 mrad. HAADF, LAADF & BF STEM imaging are performed with detector collection angle as 63-200, 15-59 & 0-13 mrad respectively. STEM-EDS elemental maps are formed after background subtraction. To reduce noise, each elemental map is convolved with a two-dimensional Gaussian (3 pixel standard deviation).

**Supplementary Text and Figures**

*Device simulation results*

Simulations (**Fig. S1**) show that the sheath of ionized DLS-active point defects near the $MoO_3$/CdS junction is wider in the ON state than in the OFF state. These simulations predict a substantial junction capacitance (*e.g.* 135 nF/cm$^2$ in the OFF state) and capacitance switching effect, to accompany the resistive switching. Our experimental capacitance measurements suggest that the devices do have very large capacitance, on the order of $10^{-6}$ F/cm$^2$, but the measurements are not accurate due to the small device resistance. The internal impedance of the LCR meter used is 2 kΩ, which is very large compared to the device under test.

Although our device simulations allow us to predictively design devices, and provide insight into the operating mechanism, accurate simulations of *I-V* data remain a substantial challenge that requires modeling bistable and history-dependent defect configurations and charge sates, substantial internal electric fields, tunneling, and recombination currents.

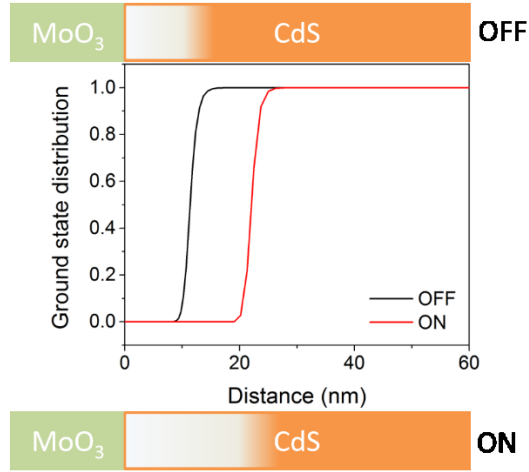

**Figure S1:** Simulated distribution of bistable, DLS-active defects as a function of distance from the $MoO_3$/CdS interface in equilibrium-ON devices, determined for ON (bias -0.4 V) and OFF (bias 0.4 V) states. The illustrations show how the sheath (white-to-light orange) of ionized DLS-active defects is narrow in the OFF state, and wide in the ON state.

*Device testing results: Area-dependence*

A characteristic property of resistive switching based on localized filaments is a weak dependence of resistance on device area [22]. We measured the *I-V* curves of three equilibrium-ON devices with different area. As we show in F**ig. S2**, both the ON and OFF state resistances scale inversely with device area for devices smaller than 1 mm$^2$, as expected for a laterally-uniform switching mechanism, and which is inconsistent with filament-based switching. The background series resistance of our devices is approximately 3.6 Ω, which comes from the silver substrate and the measurement apparatus. This is large compared to the ON state resistance of devices larger than 1 mm$^2$ (which is under 1 Ω). Therefore, the ON-state area dependence of the device resistance is weak for larger devices.

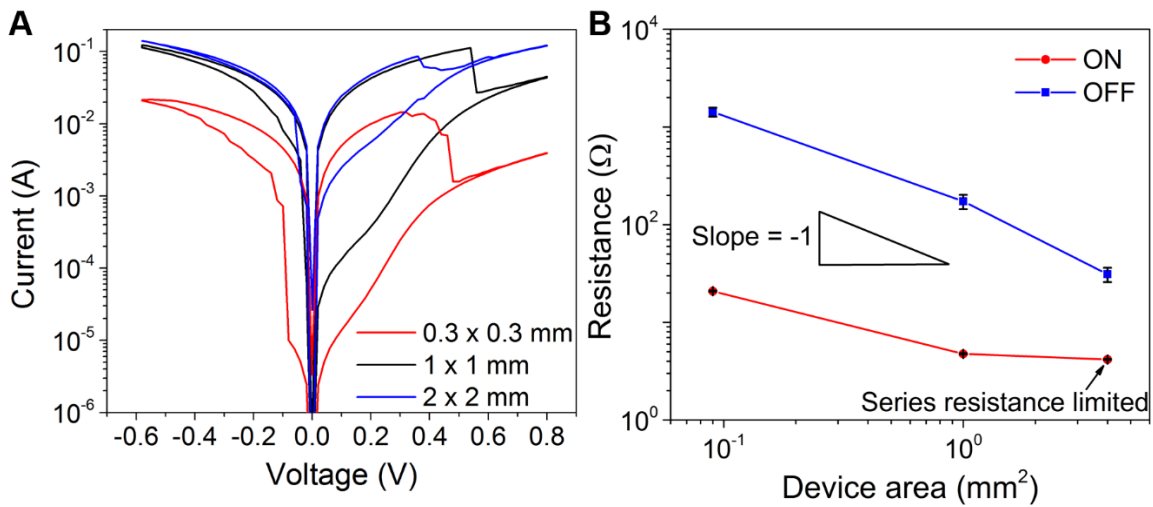

**Figure S2:** Size-dependence of equilibrium-ON device test results. (**A**) I-V characteristics of resistive switching devices with different sizes. (**B**) Device resistances extracted at 0.3 V for both ON and OFF states.

*Device testing results: The role of metal layers*

In **Fig. S3** we show additional device test results that illuminate the role of the metal layers in the device performance. As an additional test for the possible role of Ag filaments, beyond those presented in the main text, we replace Au with Al in the top contact. Al has a much lower work function, and as a result the $MoO_3$/CdS junction is insufficiently hole-injecting to operate as an equilibrium-ON DLS device. However, the choice of top contact metal should not affect the growth of Ag filaments from the bottom contact into the CdS layer. This device is in the low-conductivity (OFF) state as-fabricated, and shows no resistive switching within a ±1 V window. This further demonstrates the growth of Ag filaments is unlikely to be important to the equilibrium-ON device functionality. By way of contrast, when we replace Ag by Au as the bottom contact metal, the devices show evidence of competing mechanisms. In **Fig. S3** we show an *I-V* test of a Au/CdS/$MoO_3$/Ag device. This device is equilibrium-ON, and does show switching in the same sense as our standard devices, but the *I-V* curves are messy and the switching behavior is non-repeatable. We suggest that in this case the Au atoms are mobile throughout the CdS active layer, driven by the known affinity of Au and S.

We also fabricated and tested Hf/CdS/Al devices. Based on the similar and low work functions of Hf and Al we expect little-to-no built-in voltage in this device and nearly-ohmic contacts. The *I-V* data (**Fig. S4**) shows ohmic transport and no sign of resistive switching even out to ±4 V (implying an electric field in the CdS layer of approximately 0.7 MV/cm).

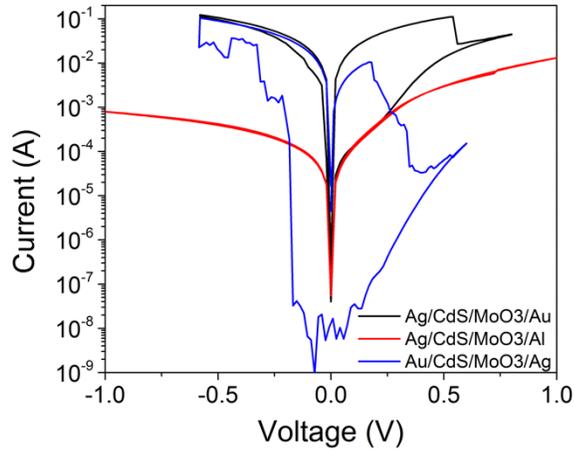

**Figure S3:** *I-V* characteristics of alternative devices further demonstrating the role of metal layers in the performance of equilibrium-ON devices.

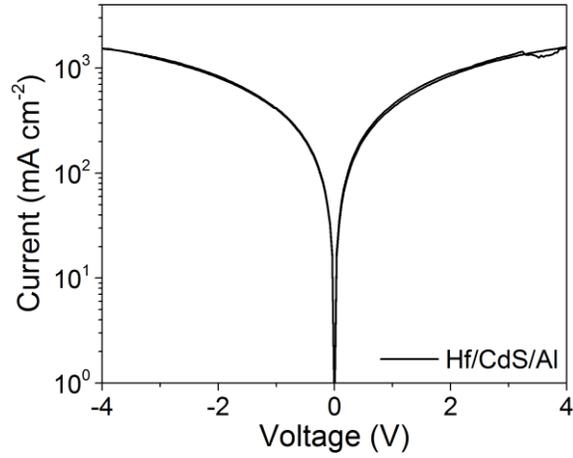

**Figure S4:** *I-V* characteristics of a Hf/CdS/Al device, showing no resistive switching effect even to $\pm 4$ V.